\documentclass[aps,prl,twocolumn,showpacs,preprintnumbers,superscriptaddress]{revtex4}

\usepackage{graphicx}
\usepackage{dcolumn}
\usepackage{bm}

\begin{document}


\title{Steplike Lattice Deformation  of  Single Crystalline (La$_{0.4}$Pr$_{0.6}$)$_{1.2}$Sr$_{1.8}$Mn$_{2}$O$_{7}$ Bilayered Manganite }

\author{M.Matsukawa} 
\email{matsukawa@iwate-u.ac.jp }
\author{Y.Yamato}
\author{T.Kumagai}
\affiliation{Department of Materials Science and Technology, Iwate University , Morioka 020-8551 , Japan }
\author{R.Suryanarayanan}
\affiliation{Laboratoire de Physico-Chimie de L'Etat Solide,CNRS,UMR8182
 Universite Paris-Sud, 91405 Orsay,France}
\author{S.Nimori}
\affiliation{National Institute for Materials Science, Tsukuba 305-0047 ,Japan} 
\author{M.Apostu}
\author{A.Revcolevschi}
\affiliation{Laboratoire de Physico-Chimie de L'Etat Solide,CNRS,UMR8182
 Universite Paris-Sud, 91405 Orsay,France}
\author{K. Koyama}
\author{ N. Kobayashi}
\affiliation{Institute for Materials Research, Tohoku University, Sendai  
980-8577, Japan}

\date{\today}

\begin{abstract}
We report  a steplike  lattice transformation of single crystalline (La$_{0.4}$Pr$_{0.6}$)$_{1.2}$Sr$_{1.8}$Mn$_{2}$O$_{7}$bilayered manganite accompanied by both magnetization and magnetoresistive jumps, and examine the ultrasharp nature of the field-induced first-order transition from a paramagnetic insulator to a ferromagnetic metal phase accompanied by a huge
decrease in resistance.
Our findings support that the abrupt magnetostriction  is closely related to an orbital frustration existing in the inhomogeneous paramagnetic insulating phase rather than  a martensitic scenario between competing two phases.

\end{abstract}

\maketitle
The lattice transformation  associated with  a field-induced  metamagnetic transition, observed in perovskite manganites and a class of rare-earth intermetallic compounds, gives rise to significant information for our understanding of
the dynamics of first-order phase transitions 
in  solid-state physics\cite{TO00}.
For example, the structural transformation of intermetallic compounds from an austinite (parent)  to a martensite 
phase upon increasing an external field is accompanied by a metamagnetic transition from an antiferromagnetic
(AFM) to a ferromagnetic (FM) state\cite{PE03}.
This type of transition contrasts with that activated  thermally where thermal fluctuations
drive the system from a metastable towards a stable state at constant external parameters
(external magnetic field, temperature, and external stress)\cite{FR01}. 
A martensite transformation is  associated with a significant change  of the unit cell from the parent to the new phase and takes place as a sequence of avalanches between metastable states due to local strain fields
stored in the lattice \cite{VI94}.

Recent experimental and theoretical studies on doped manganites exhibiting
a colossal magnetoresistance (CMR) effect have revealed the inhomogeneous coexistence of
charge-ordered  AFM insulating and  FM metallic phases in a phase-separated ground-state\cite{DA01}.
The low-temperature phase-separated state
is composed of a charge-ordered(CO)  matrix phase with a large lattice distortion due to a cooperative Jahn-Teller effect and  metallic FM clusters with a different unit cell from the former phase, indicating 
the presence of substantial strains at CO/FM interfaces. 
Several recent studies on metamagnetic transitions  of manganites  have also shown that 
ultrasharp magnetization steps appear at low temperatures.
To account for this, a model based on a martensitic scenario between competing CO/FM phases has been proposed though  questions have been  raised against this model\cite{MAH02,HA03,FI04,GHI04,LI06}. 


The purpose of this Letter is to present  a detailed investigation  of  anisotropic magnetostrictions of
 single crystalline (La$_{0.4}$Pr$_{0.6}$)$_{1.2}$Sr$_{1.8}$Mn$_{2}$O$_{7}$ , and to examine the dynamics of a
steplike first-order transition from a paramagnetic insulating (PMI) to a ferromagnetic metal(FMM) phase in CMR manganites.
Our data strongly favor an orbital frustration to account for our findings. 
For the Pr-substituted (La$_{0.4}$Pr$_{0.6}$)$_{1.2}$Sr$_{1.8}$Mn$_{2}$O$_{7}$ crystal,  a spontaneous ferromagnetic metal phase (originally present with no Pr substitution) disappears at  ground state but a field-induced PMI to FMM transformation is observed over a wide range of temperatures. A magnetic  ($H,T$) phase diagram, established from  magnetic measurements, is separated into three regions labeled as PMI, FMM , and phase-separated (PS) states (see refs.\cite{MA04,MA05}). The proximity of free energies between the PMI and FMM states is of importance to stabilize the PS state. Application of an external field
destabilizes the PMI ground state and drives the transition to the FMM state through the intermediate PS state.  As seen from the magnetic phase diagram,
a stronger field is needed to induce the PMI to FMM transition at low temperatures since the thermal energy is reduced
upon lowering temperature.  This tendency  points to a thermally activated character of the metamagnetic transition.    

Single crystals of (La$_{0.4}$Pr$_{0.6}$)$_{1.2}$Sr$_{1.8}$Mn$_{2}$O$_{7}$ were grown by the floating zone method using  a mirror furnace. 
The lattice parameters as well as the chemical homogeneity of  the tetragonal crystal were reported earlier \cite{AP01}.  The dimensions of  the $z$=0.6 sample are 3.4$\times$3 mm$^2$ in the $ab$-plane and 1mm along the $c$-axis.  Measurements of magnetostriction , both in the $ab$-plane and along the $c$-axis , were done by means of a conventional strain gauge method at the Tsukuba Magnet Laboratory, the National Institute for Materials Science (NIMS) and at the High Field Laboratory for Superconducting Materials, Institute for Materials Research, Tohoku University. 
The magnetization measurements were made  by using a VSM magnetometer  at NIMS.
The sample was zero-field cooled from 300K down to low temperatures and   we then started  measuring the isothermal
magnetostriction upon increasing (or decreasing) the  applied fields, parallel to the $c$-axis.
The normal sweep rate was set to be 0.26 T/min. 
\begin{figure}[ht]
\includegraphics[width=8cm,clip]{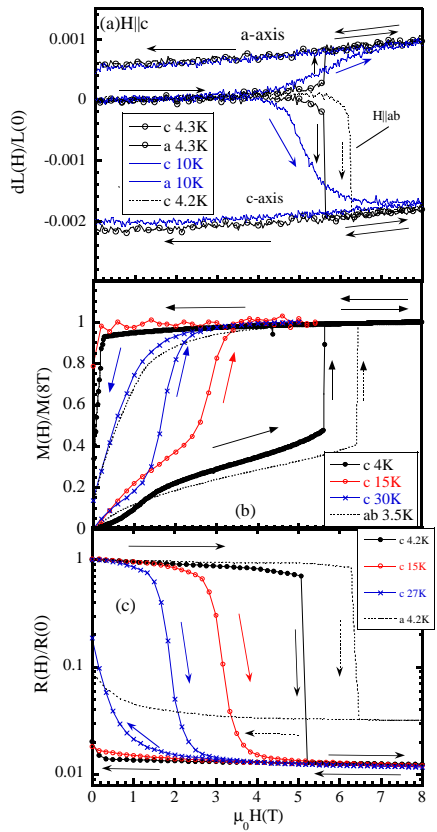}%
\caption{(color online) (a)the $c$-axis and $ab$-plane isothermal magnetostriction of a (La$_{0.4}$Pr$_{0.6}$)$_{1.2}$Sr$_{1.8}$Mn$_{2}$O$_{7}$ single crystal at $T$= 4.3 K and 10 K ($H\parallel c$). A dotted curve represents the $c$-axis data at 4.2K in the case of $H\parallel ab$.
The small magnetoresistance effect of the strain gauge itself gives a slight increase in both $dL_{c}$ and $dL_{ab}$.
(b)Field dependence of the isothermal magnetization of the same crystal at selected temperatures
($H\parallel c$). For comparison, the $ab$-plane magnetization data are given at 3.5K ($H\parallel ab$).
(c)$c$-axis  magnetoresistances of the same crystal at selected temperatures in the cases of 
$H\parallel c$. For comparison, the $\rho_{ab}$ data at 4.2 K($H\parallel ab$) are cited.
}
\label{}
\end{figure}%
Figure 1 shows the isothermal magnetostriction data of a (La$_{0.4}$Pr$_{0.6}$)$_{1.2}$Sr$_{1.8}$Mn$_{2}$O$_{7}$ single crystal at $T$= 4.3 K and 10 K. 
First of all, we emphasize qualitative differences in the magnetostriction between ultrasharp transformations 
at low-$T$ observed here and broad variations at higher-$T$ previously reported in \cite{MA04} (width$\sim $the order of 1T) .
Both  $dL_{c}$ and $dL_{ab}$ are independent of the applied field up to $\sim $5T.
However, upon further increasing the field, the $c$-axis abruptly shrinks by $\sim 0.2 \%$ at low-$T$, whereas the $a$($b$) axis expands simultaneously. The transition widths are estimated to be  within 0.1 $\%$ of the magnitude of
the critical field $H_{c}$ from the present data.
Following a steplike change of the magnetostriction  at $H_{c}$,  the values of $dL_{c}$ and $dL_{ab}$  then remain almost constant upon increasing $H$, up to 8T and decreasing it down to zero field. 
Finally, the $dL_{c}(H)$ and $dL_{ab}(H)$  curves exhibit a large hysteresis 
and the remanent striction remains very stable for about one day after switching off the field.
For comparison, the isothermal magnetization and magnetoresistance of the same composition crystal  are presented in Fig.1(b) and (c). Here, the value of $M_{c}$(8T) reaches $\sim$ 3.5$\mu_{B}$/Mn site, close to the full moment of 3.6$\mu_{B}$/Mn site, while the value of  $M_{ab}$ is $\sim$ 3.0$\mu_{B}$/Mn site, which is by about 20 $\%$ smaller than the ideal value \cite{ AP01}. 
The corresponding magnetization and magnetoresistance data also exhibit a steplike transition
at the same critical field. We note that a steplike behavior of the magnetostriction is not observed  for polycrystalline (Sm,Sr)MnO$_{3}$ manganite in spite of the appearance of both a steplike magnetization and magnetoresistance\cite{FI04}.
The steps observed in $dL(H)$,$M(H)$ and $\rho(H)$ are almost independent of the field direction.
\begin{figure}[ht]
\includegraphics[width=8cm,clip]{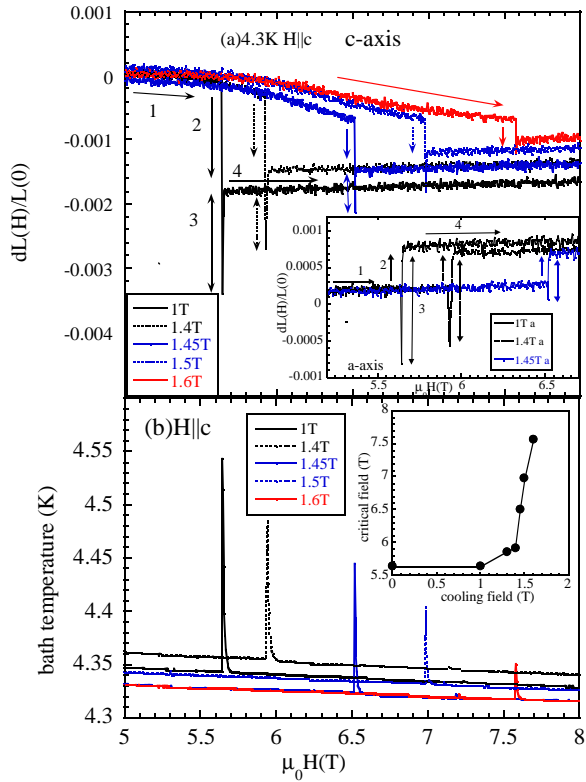}%
\caption{(color online)(a) $c$-axis magnetostriction as a function of cooling field (1,1.4,1.45,1.5 and 1.6 T).
For comparison, the $a$-axis data are displayed in the inset of (a).
First, warming the sample from 4.3K up to 200K in the absence of field and then cooling it from 200K down to 4.3K in 
the respective field during 30 min (FC), the applied field was removed. Finally, we  start measuring the magnetostriction upon increasing the field up to 8T. For each measurement, we repeated the same procedure.
The arrows labeled 1 and 4 mark $dL_{c}(H)$, before and after
the transition, respectively. Arrow 2 represents the net lattice deformation, while Arrow 3 points to
the apparent change caused by the rise of sample temperature.  
(b) The corresponding bath temperature versus applied field.  The slight decrease in the bath temperature 
as a function of $H$ arises from the magnetoresistance effect of the cernox thermometer.
The inset of (b) represents the critical field,$H_{c}$, versus the cooling field, at 4.3K.}
\label{}
\end{figure}%
Next, we present in Fig.2 (a) the $c$-axis magnetostriction data as a function of cooling field, 
because the cooling field affects a fraction of the FMM region\cite{MAH02}.
All $dL_{c}(H)$ data show a smooth decrease and then follow a steplike drop.
The value of $dL_{c}(H)$ decreases from $\sim$0.18$\%$   at the FC-1T run down to 
$\sim$0.1$\%$   at the FC-1.6T run. 
The substantial difference in $dL_{c}(H)$ between the FC-1T and FC-1.6T runs
indicates that  the FMM phase is partially formed within the PMI matrix through each FC run, before magnetostriction measurements are done.
The remarkable jump observed in $dL_{c}(H)$ is monotonically decreased upon increasing the cooling field
from 1T up to 1.6T, accompanied by a nonlinear dependence of $H_{c}$ from 5.6T up to 7.6T.
For all measurements exhibiting the steplike transition, a divergent behavior of  bath temperature was observed at the corresponding $H_{c}$ as shown in Fig.2(b).  The larger steps in  $dL_{c}(H)$ accompany a more sudden temperature rise. In addition to it, the apparent striction in both  $dL_{c}(H)$ and  $dL_{a}(H)$  at the steplike transition suggests a rapid rise of the sample's temperature  due to the released heat \cite{GAUAGE} (see the caption of Fig.2(a)).
It has been reported from the magnetocaloric measurement on polycrystalline La$_{0.225}$Pr$_{0.4}$Ca$_{0.375}$MnO$_{3}$ that a released heat at the ultrasharp field-induced transition  gives rise to the abrupt increase of the sample temperature $\delta T$ $\sim$30K
\cite{GHI04}. 
When the cooling field exceeds  1.6 T, the steplike behavior of $dL_{c}(H)$ disappears.

\begin{figure}[ht]
\includegraphics[width=7cm]{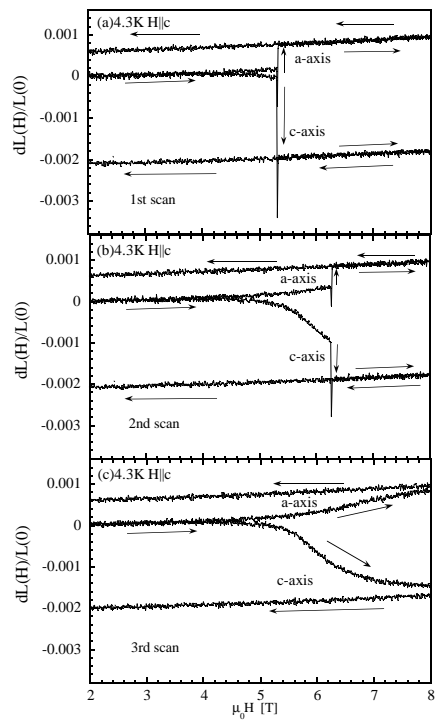}%
\caption{Field dependence of the $c$-axis and $a$-axis isothermal magnetostriction, $dL_{c}$ and $dL_{ab}$,
at 4.3K ($H\parallel c$). (a)1st scan,(b)2nd scan, and (c) 3rd scan.
Here, 1st scan of the applied field denotes 0T$\rightarrow$8T$\rightarrow$0T. 2nd (3rd) scan;local heating$\rightarrow$0T$\rightarrow$8T(12T)$\rightarrow$0T. 
(For details, see the text)}
\label{}
\end{figure}%
Then, after a first scan in zero field cooling, we  carried out  successive measurements of the isothermal magnetostriction.
At the beginning of further scans, we reset the sample from the remanent FMM to the PMI virgin state by a local heating
(see ref.\cite{LOCAL}).
In the second scan ( Fig.3(b)), once  $dL(H)$  shows a
continuous variation with the IM transition , near 6.3T,  a sudden jump along both the $c$ axis and the $a$ axis
is also observed.  However, the third scan data show no discontinuous step, similar to the normal $dL(H)$ curves at
higher temperatures.  
The critical field of the second scan is increased up to 6.3T from 5.3T in the first 
scan. The qualitative differences in the $dL(H)$ curves among the three scans probably arise from a training effect
due to the local heating from the strain gauges attached on the sample.
Such a local heating erases the memories experienced by the sample, removes frustrations existing within the PMI matrix
and stabilizes the PMI state, resulting in a widely broad transition, as seen in the third scan.
\begin{figure}[ht]
\includegraphics[width=5cm]{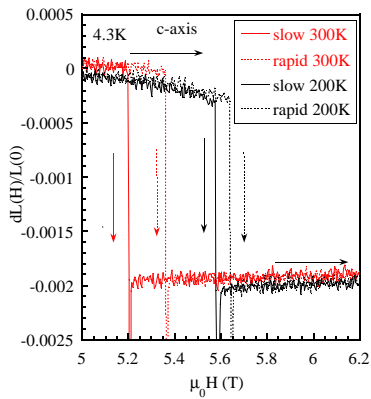}%
\caption{(color online)Cooling rate dependence of $dL(H)$ from  200 and 300 K down to 4.3K.
In the case of 200K, rapid and slow cooling rates are set at 30 min and 12 hours.
At 300K, rapid and slow cooling rates in 20 min and 15 hours, respectively.}
\label{}
\end{figure}%
For comparison, we examine how the cooling rate  from  high temperatures down to 4.3K affect the critical field.
In both rapidly and slowly cooling procedures, the step phenomena appear (Fig.4), and we get the comparable values of $H_{c}$=5.36T (rapid) and 5.2T (slow) in the case of 300K. In a similar way, we get $H_{c}$=5.63 T and 5.58 T ,for rapid and slow cooling rates from 200K,respectively. 
We found out  that $H_{c}$ is almost independent of the cooling rate and is suppressed in the cooling process  from  a higher temperature of 300K. 
In addition, checking the sweep rate dependence of $H_{c}$,
we get 5.6 T in both 0.52 and 0.26T/min and 5.8 T at the slow rate of 0.052T/min (not shown here).

Finally, we discuss now the origin of the magnetostrictive jump observed here.
The parent  bilayered manganite  La$_{1.2}$Sr$_{1.8}$Mn$_{2}$O$_{7}$ exhibits a PMI to FMM transition around $T_{c}\sim$120K  and an associated CMR effect \cite{MO96}. In the half-doped bilayered manganite of La$_{2-2x}$Sr$_{1+2x}$Mn$_{2}$O$_{7}$($x$=0.5), near the $x$=0.4 crystal,
a long-range orbital- and charge-ordered (OO/CO) state  is established,accompanied by a CE-type antiferromagnetic structure\cite{AR00}. Neutron scattering measurements on a bilayered manganite near optimal doping show that CE-type OO/CO clusters (short-range polaron correlations) freeze upon cooling to $T^{*}\approx 310K $ ($T^{*}$:glass transition temperature), as a magnetic frustration between FM and AFM interactions leads to a spin-glass state \cite{AR02}.This finding  strongly suggests the existence of an orbital frustration in the PMI phase, which prevents the formation of the CO phase.
Thus, we believe that an orbital frustration exists within the PMI state of the present Pr-substituted  sample.
Once this type of frustration is initially formed at high temperatures within the sample, one expects that it survives down to low temperatures. 
In the successive isothermal scans (Fig.3),  the local heating process relaxes the frustration
existing within the matrix of the sample, giving a standard broad transition.
In other words, in the first scan, the frustration existing  probably enhances the instability of the metastable state of the free energy under the magnetic field, resulting in a steplike behavior. These findings are consistent with the fact that  the critical field is sensitive to the thermal profile undergone by the sample from high temperatures.
This scenario also accounts for the cooling field dependence of $H_{c}$  observed here and reported in magnetization steps \cite{MAH02,LI06}. The field-cooled run suppresses the CO clusters but increases the FM clusters.  Accordingly,
the reduced frustration affects a steplike transition, increasing the value of $H_{c}$, which is consistent with 
the positive dependence of $H_{c}$ on the cooling field.  
The nucleation and growth of FMM domains due to an external field reduce the magnetic energy by the Zeeman energy, 
while they increase the elastic energy arising from strains stored at the interface of two-phase domains.
It is true that this competition in free energy has some relation with the transition, but it never reproduces
quite different isothermal behaviors under the same external field scans.   

In summary, we observed a steplike  magnetostriction in a single crystalline (La$_{0.4}$Pr$_{0.6}$)$_{1.2}$Sr$_{1.8}$Mn$_{2}$O$_{7}$ sample, accompanied  by steps in both  magnetization and magnetoresistance.
Our data suggest that the abrupt lattice transformation reported here cannot be described by a martensitic transformation but  is closely related to an orbital frustration existing in the inhomogeneous PMI phase.  
This work was supported by a Grant-in-Aid for Scientific Research from Japan Society of the Promotion of Science.


\begin{thebibliography}{30}
\bibitem{GAUAGE}Upon decreasing $T$ from 30K down to 4K,  the resistance of the commercial strain gauges  used (KYOWA) for both $ab$-plane and $c$-axis low-temperature measurements  increases by about 10$^{-3}$. 
At low temperatures, the thermal contraction of solids is negligible (at most  $dL/L$ $\sim$ the order of  10$^{-5}$) in comparison with the apparent change of the strain gauge itself. Accordingly, we conclude that the apparent striction associated with the step transition is caused by a heat generation within the crystal.
\bibitem{LOCAL}For the local heating,  a current of 15mA was applied to a series of  two gauges during 10 seconds so that
the gauge resistance  returned rapidly to an initial value before the application of field, corresponding to
the PMI virgin state.
After checking the stability of the bath temperature, we then restarted measuring $dL(H)$ . 
For each measurement, the excited current of the gauge was set to be 1 mA.
\end{thebibliography}
\end{document}